\documentclass{appolb}

\usepackage{graphicx}
\usepackage{xcolor}
\usepackage{xspace} % spaces after comands
\usepackage{subfigure}
\usepackage{wrapfig}
\usepackage[utf8]{inputenc}
\usepackage[OT1]{fontenc}

\newcommand{\Fred}{\textsc{Fred}\xspace}

\begin{document}
\title{Investigations on physical and biological range uncertainties in Krakow proton beam therapy centre
\thanks{Presented at 3rd Jagiellonian Symposium on Fundamental and Applied Subatomic Physics, June 23 – 28, 2019 in Collegium Maius, Kraków, Poland}%
}
\author{A.~Rucinski$^a$, J.~Baran$^a$, G.~Battistoni$^{bc}$, A.~Chrostowska$^d$, M.~Durante$^{ef}$, J.~Gajewski$^a$, M.~Garbacz$^a$, K.~Kisielewicz$^d$, N.~Krah$^g$, V.~Patera$^h$, M.~Pawlik-Niedźwiecka$^{ai}$, I.~Rinaldi$^j$, B.~Rozwadowska-Bogusz$^d$, E.~Scifoni$^c$, A.~Skrzypek$^a$, F.~Tommasino$^{ck}$, A.~Schiavi$^h$, P.~Moskal$^i$
\address{(a) Institute of Nuclear Physics Polish Academy of Sciences, Kraków, Poland \\
(b) INFN, Milan, Italy
(c) TIFPA, Trento, Italy
(d) Maria Sklodowska-Curie Memorial Cancer Center and Institute of Oncology Cracow Branch, Kraków, Poland 
(e) Biophysics Department, GSI Helmholtzzentrum f\"ur Schwerionenforschung, Darmstadt, Germany 
(f) Technische Universit\"at Darmstadt, Institut f\"ur Festk\"orperphysik, Darmstadt, Germany 
(g) University Lyon, CNRS, CREATIS UMR 5220, Centre Leon Berard, Lyon, France \\
(h) Sapienza University of Rome, Rome, Italy 
(i) Institute of Physics, Jagiellonian University, Kraków, Poland 
(j) ZonPCT/Maastro clinic, Maastricht, The Netherlands 
(k) University of Trento, Trento, Italy
}
}
\maketitle
\begin{abstract}
Physical and biological range uncertainties limit the clinical potential of Proton Beam Therapy (PBT). 
In this proceedings, we report on two research projects, which we are conducting in parallel and which both tackle the problem of range uncertainties. One aims at developing software tools and the other at developing detector instrumentation. 
%Our research activities aim at developing software tools and detector instrumentation to tackle the problem of the beam range uncertainties in the clinic. 
Regarding the first, we report on our development and pre-clinical application of a GPU-accelerated Monte Carlo (MC) simulation toolkit \Fred. 
%The MC based recalculation of patient treatment plans with variable radiobiological effectiveness is an essential input for medical doctors and physicists and can support PBT treatment planning and quality assurance. 
Concerning the letter, we report on our investigations of plastic scintillator based PET detectors for particle therapy delivery monitoring. We study the feasibility of Jagiellonian-PET detector technology for proton beam therapy range monitoring by means of MC simulations of the $\beta^+$ activity induced in a phantom by proton beams and present preliminary results of PET image reconstruction. Using a GPU-accelerated Monte Carlo simulation toolkit \Fred and plastic scintillator based PET detectors we aim to improve patient treatment quality with protons.

\end{abstract}
%\PACS{PACS numbers come here}
  
\section{Introduction}

%Proton Beam Therapy (PBT) is a rapidly growing technique for tumor radiation therapy. 
The increasing numbers of proton facilities and successful proton treatments \cite{Durante2017} indicate that the relevance of  Proton Beam Therapy (PBT) as technique for tumor radiation therapy is a rapidly growing. Kraków proton facility is in clinical operation since Oct.~2016 and more than 10 patients a day are currently treated. 

Range uncertainties, i.e.\ uncertainty of the distance that protons travel inside patient body, currently represent one of the biggest caveats for the exploitation of the full potential of proton therapy treatments \cite{Paganetti2012b}. Proton beams range is particularly affected by biological and physical uncertainties in a heterogeneous patient body. Therefore, to assure target coverage, medical physicists currently apply up to about 1\,cm safety margins around the tumor volume, which lead to the unwanted irradiation of the healthy tissues surrounding the tumor \cite{Paganetti2012b}.

The biological dose, $D_{RBE}$ expressed in $Gy(RBE)$, delivered to the patient is the actual quality of clinical interest. It is calculated as $D_{RBE}\!=\!D\!\times\!RBE$, where $D$ is the physical dose expressed in $Gy$ and $RBE$ is the Relative Biological Effectiveness.
By definition, in conventional therapy with photons $RBE\!=\!1.0$ therefore, physical and biological doses are equal and correlated with clinical response. Protons have an increased biological effectiveness compared to photons, i.e.\ RBE is larger than one. Currently in clinical routine, the RBE of protons is assumed to be constant and equal to 1.1 \cite{Paganetti2002b}. This convention neglects complex, often nonlinear dependency of the RBE on such parameters as penetration depth, Linear Energy Transfer (LET), dose, fractionation scheme, tissue type and endpoint, cell cycle phase or oxygenation level. These dependencies might affect the effective proton range, i.e., introduce biological range uncertainty and thus affect the dose to the surrounding tissue and Organs at Risk (OAR). Modification of proton physical dose by RBE, which is an uncertain weighting factor, makes the correlation of proton biological dose and clinical effect of tumor irradiation uncertain, and unification of clinical studies comparing the effectiveness of different radiation modalities challenging.

An improvement resulting from correctly applying radiobiological assumptions in PBT could be achieved only under the condition that the physical dose is accurately delivered to the patient. In fact, physical range uncertainties, occurring due to patient mispositioning or Computed Tomography (CT) scanner calibration introducing Hounsfield Unit (HU) to stopping power conversion, could cause differences between treatment plan and treatment delivery. 
%For these reasons, strategies are being developed to control/monitor physical range uncertainties. 
Monte Carlo simulations and range monitoring methods are essential in PBT to guarantee that the physical dose is accurately delivered and therefore to reduce the biological and physical range uncertainties in a patient body.
%and to fully exploit the advantages of PBT.

\begin{figure}[th!] 
\centering{}
\includegraphics[width=1\textwidth]{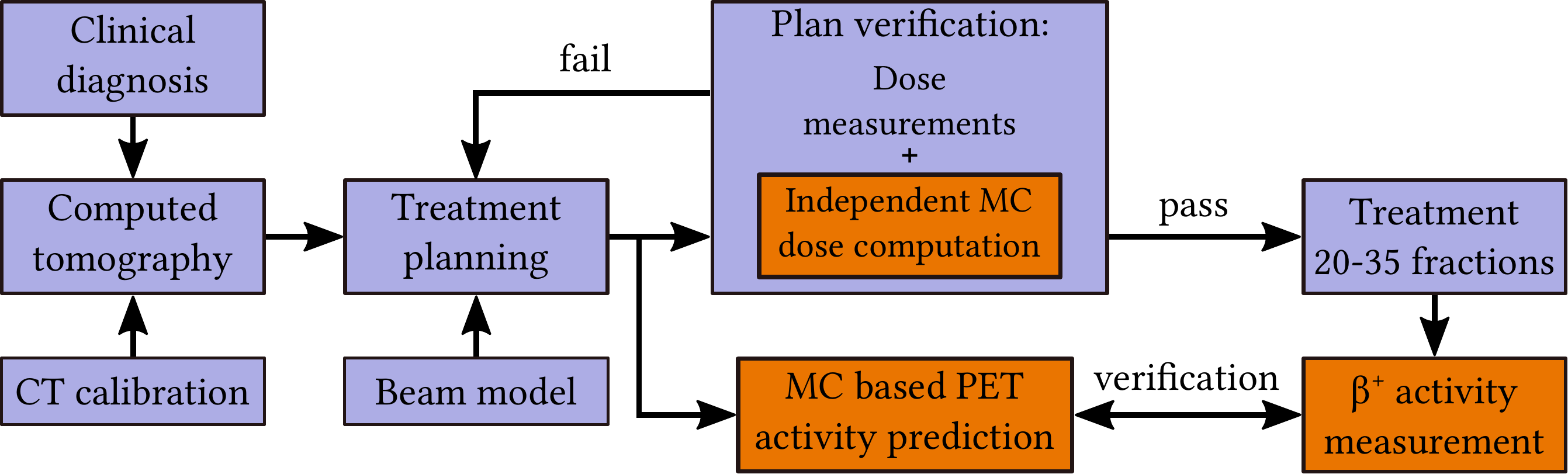} 
\caption{Diagram of standard patient treatment protocol (blue colors) along with activities of our group (orange color) aiming to improve PBT quality. We incorporate advanced nuclear physics computational and experimental methods to predict physical and biological dose in patient more accurately and to monitor proton beam range in patient by means of PET imaging.}
\label{fig:workflowChart}
\end{figure}

Fig.~\ref{fig:workflowChart} illustrates which role could play each of the two topics reported in this manuscript in the clinical treatment workflow.
%This manuscript is an overview of investigations conducted by our group within national and international collaborations aiming to tackle the problem of physical and biological range uncertainties in proton beam therapy. The activities of our group in the context of standard clinical PBT protocol is presented in fig.~\ref{fig:workflowChart}.
%\textcolor{red}{Here we can put the scheme and add two sentences of description how we are going to address range problem in context of the PBT treatment protocol.}

\section{Monte Carlo simulations to address physical and biological range uncertainties of proton beams}

It is recognized that Monte Carlo (MC) methods can offer improved dose calculation accuracy in heterogeneous media and therefore predict more accurately therapeutic dose distributions in patients compared to the analytical algorithms that are typically employed in the Treatment Planning Systems (TPS) used in clinical routine \cite{Paganetti2012b}. 
%In the context of the presented research projects, 
Nowadays, the use of a variable RBE is being discussed among the PBT scientific community.
We perform biological dose calculations with variable RBE and investigate biological range uncertainties by means of MC simulations of patient CT images exploiting the MC dose calculation tool \Fred \cite{Schiavi2017, Garbacz2019}. \Fred offers a unique combination of features: accuracy of a MC code including biological dose computation, flexibility of a research tool, and high dose calculation speed due to GPU-acceleration. These characteristics are impossible to achieve with the currently available commercial TPS and general purpose MC codes like Geant4/FLUKA \cite{Agostinelli2003a, Battistoni2007}.

\begin{figure}[ht!]
\centering{} 
\includegraphics[width=0.95\textwidth]{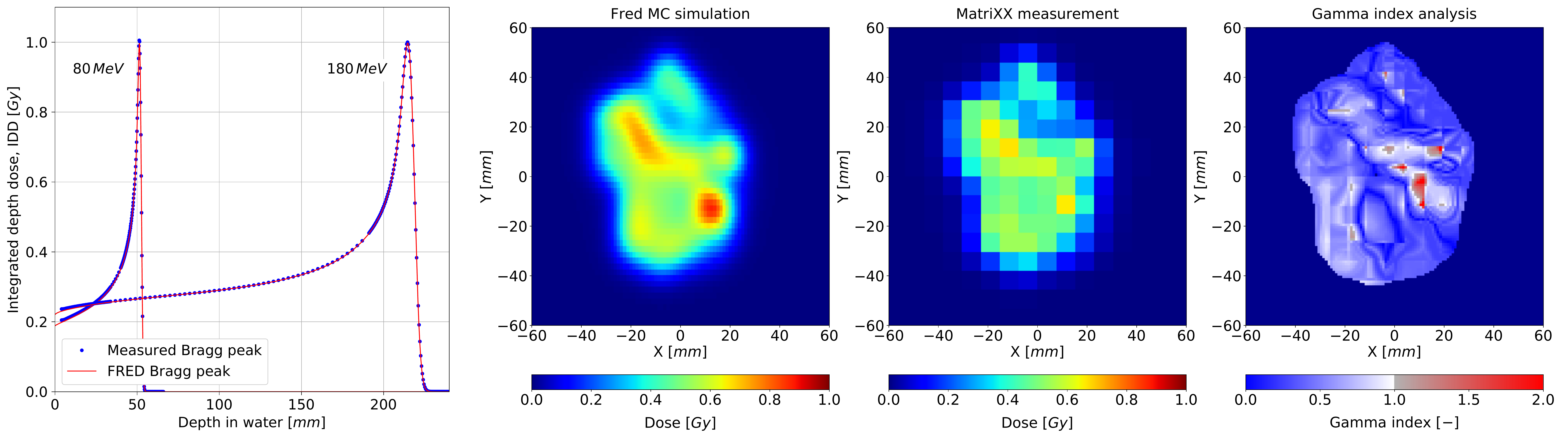} 
\caption{Integrated depth dose distributions of proton pencil beams simulated in \Fred and obtained experimentally during the facility commissioning (left), an example of transversal 2D dose distribution in isocentre plane obtained from \Fred MC simulations (middle left), measured with MatriXX detector in water (middle right) and a GI map computed from simulation and measurement using GI (3\%/2\,mm) method (right). GI passing rate is $98.64\%$.}
\label{fig:GIforLayer_and_BP}
\end{figure}

%\newpage
We have successfully implemented in \Fred the proton beam model used clinically for patient treatment in the Kraków facility. The Integrated Depth Dose profiles (IDD) of single proton pencil beams in water simulated with \Fred MC code for different energies are in excellent agreement with the IDDs obtained during the commissioning measurements (see fig.~\ref{fig:GIforLayer_and_BP} left), showing differences of less than 2\%. 

%The lateral profiles of the proton pencil beams in air were measured and fit with Gaussian in X and Y plane. The so obtained sigma-x and sigma-y parameters were fit as a function of measurement position with respect to the isocentre using an emittance model. 
% \begin{wrapfigure}{r}{0.40\textwidth}
%   \begin{center}
%     \includegraphics[width=0.40\textwidth]{braggPeak_E80_180.pdf}
%   \end{center}
%   \caption{{\footnotesize Integrated depth dose distributions of proton pencil beams simulated in \Fred and obtained experimentally during the facility commissioning.}}
%   \label{fig:BPandEmit}
% \end{wrapfigure}
\noindent The emittance model, describing lateral beam propagation in air, was implemented in \Fred, in accordance with the clinical system. The longitudinal and lateral pencil beam shapes are modelled in \Fred MC with a submillimeter precision. 

We validated the beam model experimentally using the transversal patient treatment plan verification measurements. Such measurements are routinely performed by medical physicists for treatment plan quality assurance with an array of 1020 ionization chambers (MatriXX IBA) placed in a water phantom. A transversal dose plane extracted from \Fred MC and the dose distribution measured with the MatriXX detector at the same depth in water, as well as the Gamma Index (GI) map obtained from GI test are presented in fig.~\ref{fig:GIforLayer_and_BP}. The GI passing rate ($3\%/2$\,mm criteria) greater than 98\% was obtained for 182 dose plane measurements for 10 patients. Based on these results, we can assure that dose distributions of clinical treatment plans can be recalculated accurately.  

%and considered to significantly improve the treatment quality. 
%The current knowledge is in favor of the use of a variable RBE, which is nowadays being discussed among the PBT scientific community. In addition to physical dose calculation, \Fred offers biological dose calculation with variable RBE. %The clinically relevant biological dose calculation is possible only under the condition that the physical dose calculation is precisely reproduced, as shown in Fig 3. 
We are currently performing treatment planning studies to quantify the biological range uncertainties exploring various biological models with variable RBE and clinical data of patients treated in Kraków.

\begin{figure}[hb!]
\centering{} 
\includegraphics[width=0.24\textwidth]{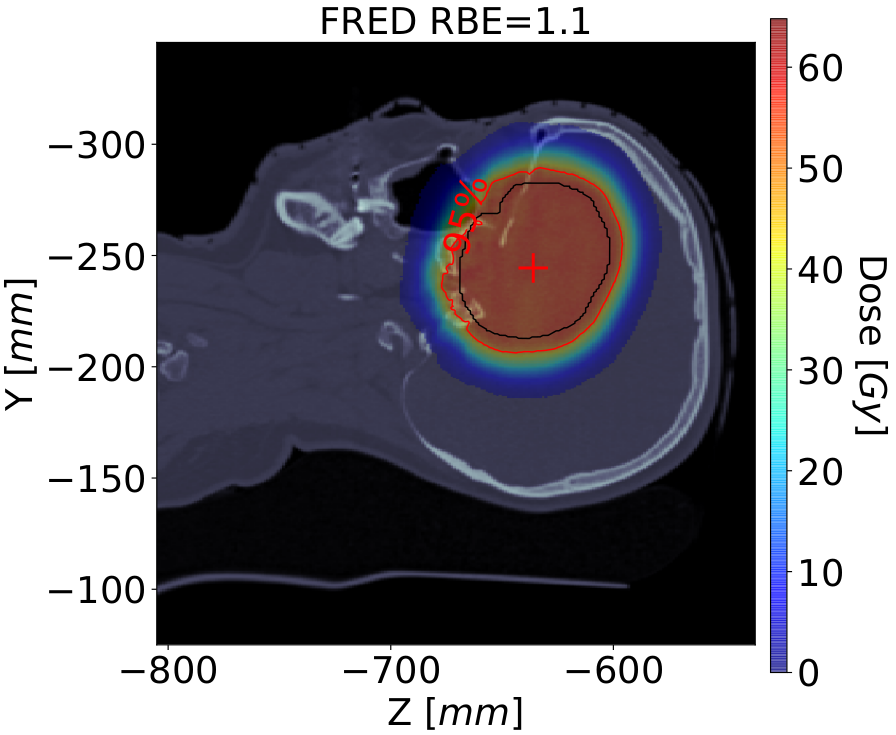} 
\includegraphics[width=0.24\textwidth]{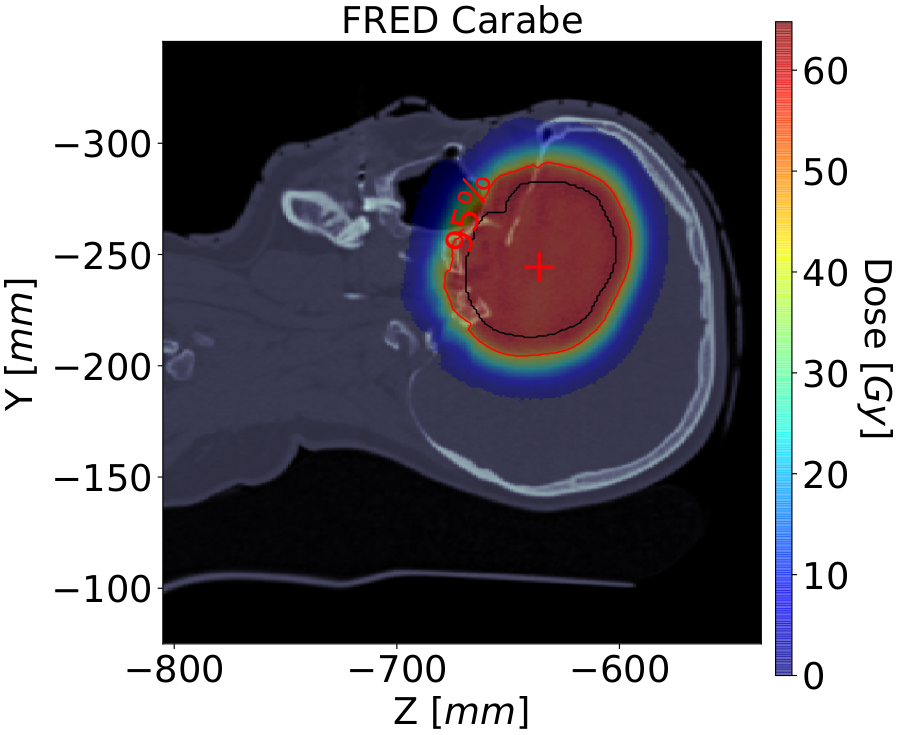} 
\includegraphics[width=0.24\textwidth]{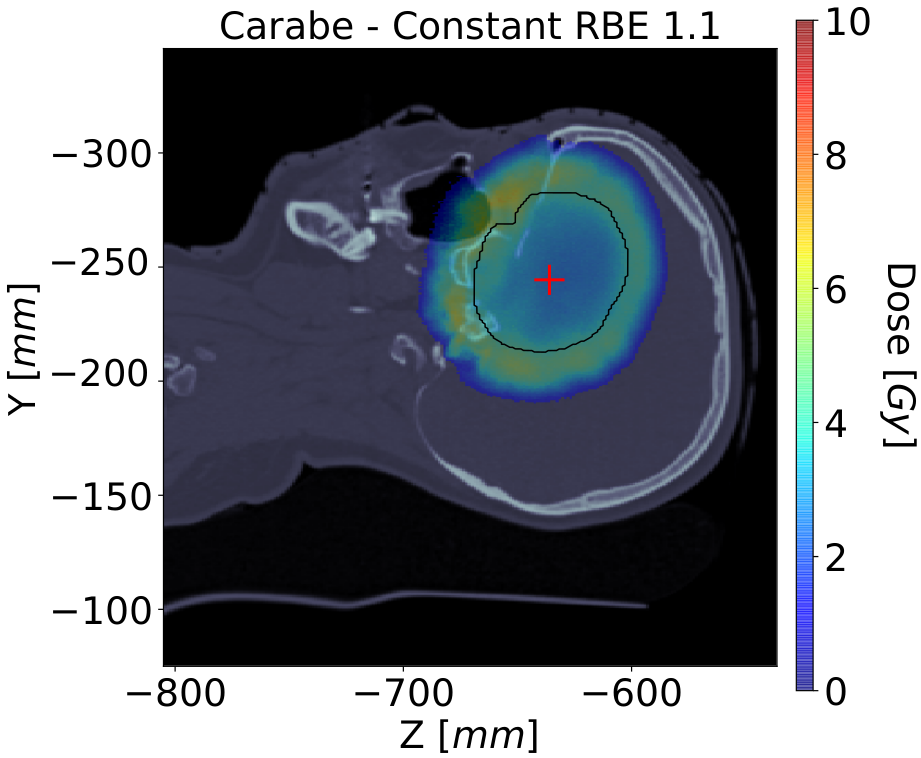} 
\includegraphics[width=0.25\textwidth]{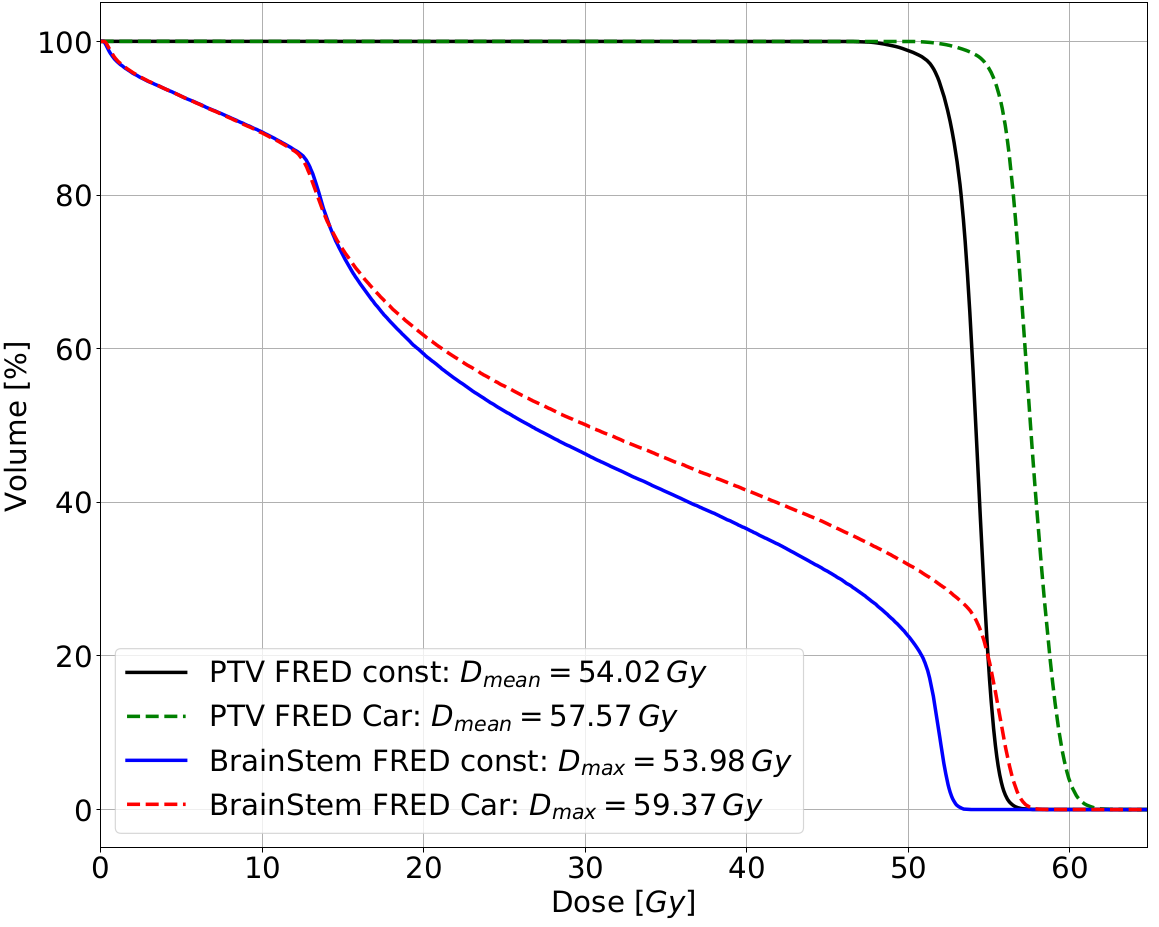} 
\caption{Left: the RBE-weighted dose with constant RBE=1.1, middle-left: the  RBE-weighted dose with variable RBE (Carabe model \cite{Carabefernandez2012}), middle-right: dose difference between constant and variable RBE. DVH for constant and variable RBE for PTV and brain stem.}
\label{fig:patientResults}
\end{figure}

An example of a selected Head and Neck (H\&N) patient is presented in fig.~\ref{fig:patientResults}, where the radiobiological dose distributions computed with constant and variable RBE using \Fred as well as corresponding Dose Volume Histograms (DVHs) are shown. The MC calculation time for this case was about 10\,min (9$\cdot$10$^8$ primary protons, 2$\cdot$10$^6$\,$\frac{p^+}{s}$ tracking rate). The mean dose to Planning Target Volume (PTV), calculated with \Fred using variable RBE model proposed by Carabe \cite{Carabefernandez2012}, is increased with respect to the clinically applied constant RBE=1.1 assumption of about 3\,Gy(RBE), whereas the maximum dose to the brain stem OAR increases by about 4\,Gy(RBE). Our results show that incorporation of the variable RBE model in patient dose calculation increases dose in PTV and OAR with respect to constant RBE assumption. This is especially important when OARs are very close to or overlap with the PTV as it occurs frequently in H\&N patients. Considering variable RBE hypothesis in the proton therapy clinic could be essential as increased biological dose deposited to OARs that exceeds clinical dose constraints can be potentially associated with increased normal tissue complication probability an therefore an increased risk of necrosis or secondary cancer.

The use of the fast MC dose computation tool \Fred for physical and biological dose recalculation of patient treatment plans (retrospectively and prospectively) can provide additional clinical information for medical physicists and medical doctors and can potentially prevent inaccuracies in patient treatment.

\section{J-PET detector to address physical range uncertainties of proton beams}
The proton interactions with patient tissues allow range monitoring during or just after the treatment detecting emitted secondary radiation. Tracking of prompt-gamma, PET-gamma and secondary protons and neutrons are examples \cite{Bauer2013a,Krimmer2018a,Moskal2019}. Prototype systems for prompt-gamma and PET-gamma range monitoring were tested clinically and obtained satisfying precision of Bragg-peak position monitoring on-line in the PBT treatment room \cite{Ferrero2018,Richter2016a,Hueso-Gonzalez2018}. At the Jagiellonian University in Kraków, a novel solution for diagnostic PET imaging, Jagiellonian-PET (J-PET) is being developed. %The possible application of J-PET detector in PBT is investigated in the frame of the project we present in this manuscript.
%\cite{Muraro2016}

A single detection unit of the J-PET scanner \cite{Kowalski2018} consists of a 50\,cm long and 6$\times$24\,mm$^2$ intersection size scintillator strip. The light pulses produced in the strip by 511\,keV back-to-back photons propagate to its edges where they are converted into electrical signals by photomultipliers (PMT). The interaction position of the photon with the detector is estimated from the time difference between the PMT signals located at the ends of the strip. A J-PET module consists of 13 scintillator strips read-out through a single front-end electronics and a FPGA-based DAQ system. A modular, lightweight and portable design of J-PET enables flexibility in detector configuration and easy installation. Increasing the number of J-PET detector layers increases the detection efficiency of the system. 

%We perform comprehensive MC simulations using GATE toolkit \textcolor{green}{[ref]} and reconstructed 3D images of $\beta^+$ activity distributions aiming at characterization of J-PET sensitivity for proton beam range detection. We investigate various geometrical configurations of J-PET modules that can be possibly applied for PET-based in-room PBT range monitoring. We simulate and perform full reconstruction of PET images, considering as the signal number of reconstructed PET coincidences per primary proton impinging PMMA phantom (fig.\,\ref{fig:jpet_barrel}).
%The reconstruction is performed using CASTOR software toolkit \textcolor{green}{[ref]} and takes into account random, scatter, attenuation and normalization corrections. 

\begin{figure}[ht!]
\centering{} 
\includegraphics[width=0.49\textwidth]{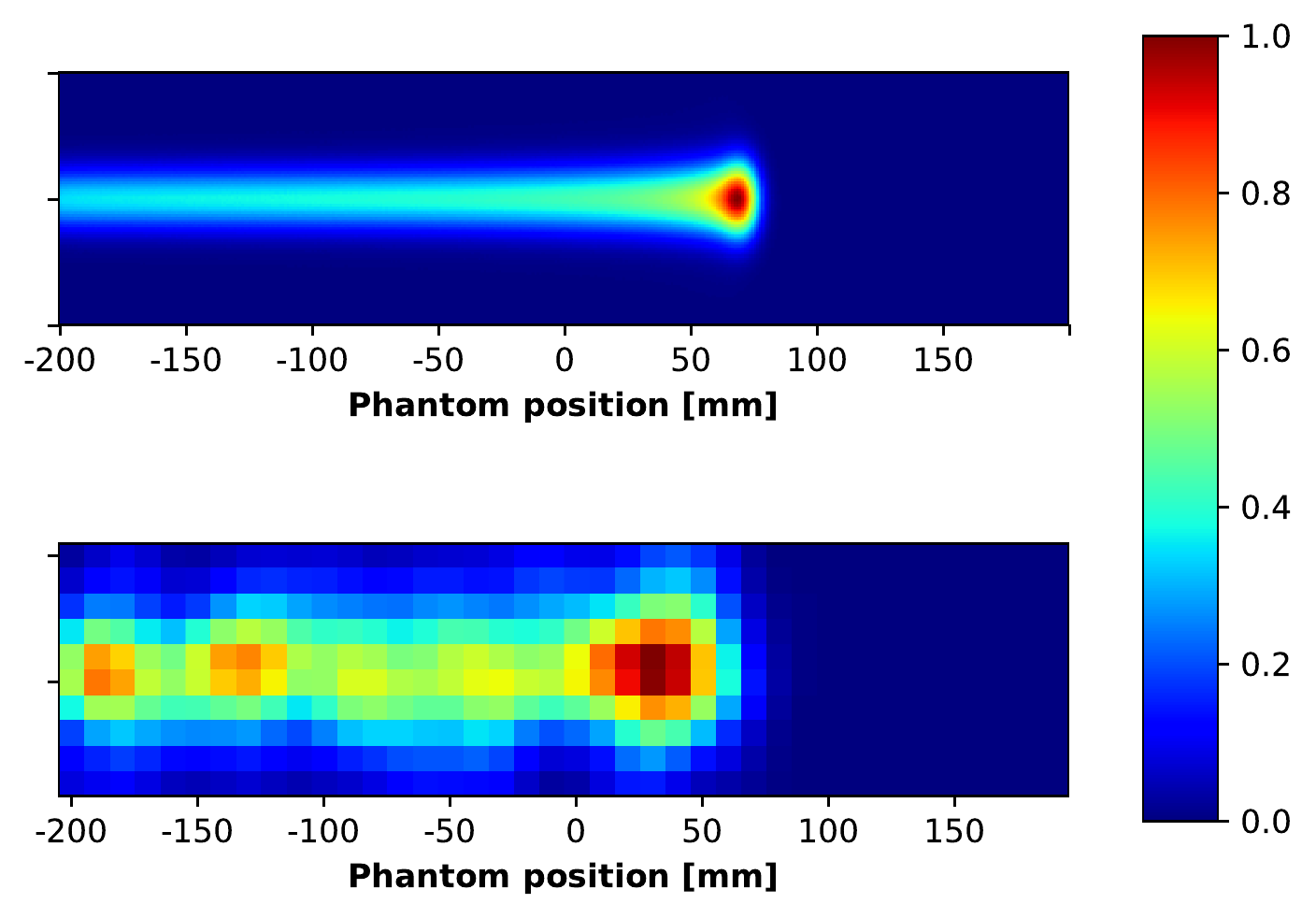} \includegraphics[width=0.49\textwidth]{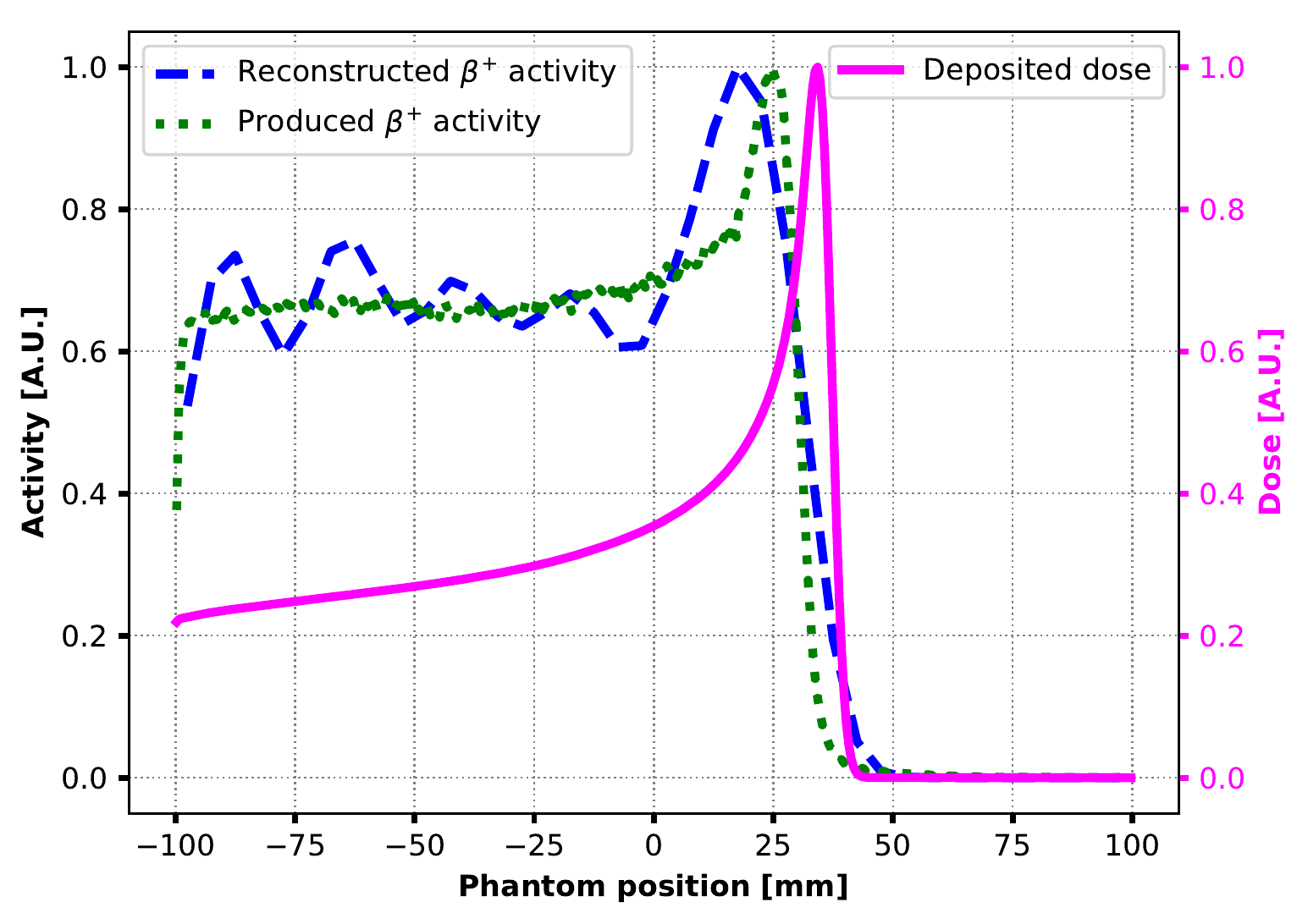} 
\caption{The results of MC simulations. Top left: 2D distribution of dose deposited by 150\,MeV proton beam in PMMA phantom; bottom left: $\beta^+$ activity distribution detected in J-PET and reconstructed with CASTOR software; right: dose, $\beta^+$ production in PMMA, reconstructed signal from $\beta^+$ activity detected with J-PET.}
\label{fig:jpet_signal}
\end{figure}

We performed comprehensive MC simulations using the GATE toolkit \cite{Sarrut2014} and reconstructions of 3D $\beta^+$ activity distributions using the CASTOR software \cite{Merlin2018}. The aim was to characterize the sensitivity of the J-PET for proton beam range detection. We investigated single and multi-layer cylindrical and dual-head configurations of the J-PET modules that can be possibly applied for in-room range monitoring. The list-mode TOF-MLEM reconstruction (5 iterations with 500\,ps TOF resolution without regularization), takes into account random events, scatter, attenuation and normalization corrections. Eventually, the reconstructed PET-activity profiles can be correlated with the position of dose distal fall-off (Fig.\,\ref{fig:jpet_signal},\,right) and used for proton beam range monitoring.
% list-mode TOF-MLEM, 5 iterations with 500 ps TOF resolution without any regularization,
%We simulate and perform full reconstruction of PET images, considering as the signal number of reconstructed PET coincidences per primary proton impinging PMMA phantom (fig.\,\ref{fig:jpet_barrel}).

%CASTOR toolkit is an open source PET image reconstruction software that fulfils J-PET image reconstruction requirements, i.e., long axial field of view, multi-layer configuration, non-cylindrical geometry and accounts for time of flight information. 

\begin{wrapfigure}{r}{0.30\textwidth}
  \begin{center}
    \includegraphics[trim={3.0cm 2.0cm 3.0cm 1cm},clip,width=0.30\textwidth]{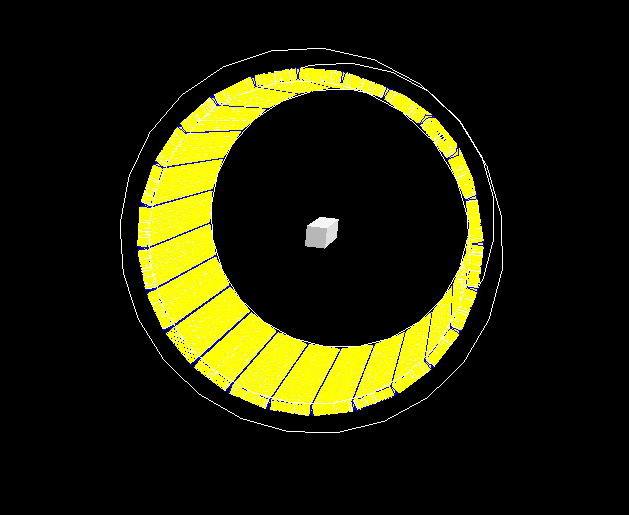}
  \end{center}
  \caption{{\footnotesize Single layer modular J-PET barrel with isocentrically positioned PMMA phantom.}}
  \label{fig:jpet_barrel}
\end{wrapfigure}
In this manuscript, we present preliminary results for one of the investigated setup configurations, i.e.\ a single layer J-PET barrel (fig.~\ref{fig:jpet_barrel}). Fig.~\ref{fig:jpet_signal} illustrates a cross-section through the centre of the beam of 3D dose distribution deposited in a 5$\times$5$\times$20\,cm$^3$ PMMA phantom by 10$^8$ protons of nominal energy 150\,MeV (top left) and the same cross-section of reconstructed 3D distribution of $\beta^+$ activity produced by the beam in the phantom (bottom left). The activity map was reconstructed in 5$\times$5$\times$5\,mm$^3$ voxel grid. The scintillator strips' detection efficiency is taken into account in MC simulations. The expected spatial and time resolutions of J-PET with wavelength-shifting strips (WLS) \cite{Smyrski2017} is taken into account through plastic length discretization used in the simulations and image reconstruction.
%According to our knowledge non of the available open-source software reconstruction tools offers J-PET specific continuous position determination along the axial direction, and signal must be discretized to virtual crystals.

The fig.~\ref{fig:jpet_signal} (right) presents the MC simulated profiles of: (i) proton dose deposition in the PMMA phantom, (ii) $\beta^+$ activity produced in the phantom and (iii) actual signal detected by J-PET barrel from $\beta^+$ activity. The results show that the J-PET detector is feasible to acquire the $\beta^+$ activity produced during proton therapy treatment and that the offline 3D reconstruction of PET activity images is possible using CASTOR toolkit. The characterization of J-PET sensitivity for proton beam range detection is currently an ongoing research activity.

\vspace{-0.8cm}
\section{Summary}
\vspace{-0.1cm}
Within the research projects conducted in the Institute of Nuclear Physics PAN in Kraków we investigate physical and biological range uncertainties of proton beams through positron emission tomography (PET) based solutions and Monte Carlo (MC) simulations. 
%The software tools and procedures developed are currently integrated into the cancer patient treatment procedures to fully exploit the advantages of proton beams in the clinic. 
Taking advantage of the \Fred accuracy and time performance possible due to GPU acceleration we aim to improve quality assurance and treatment planing in Kraków PBT facility. A Monte Carlo study of J-PET detector feasibility performed in the frame of the project suggests that this technique might be considered as a novel  proton beam therapy range monitoring approach.

\vspace{-0.3cm}
\section{Acknowledgments}
\vspace{-0.1cm}
The authors acknowledge Dosimetry and Quality Control Laboratory of CCB for supporting project activities. The FRED Monte Carlo  project is carried out within the Reintegration programme of the Foundation for Polish Science co-financed by the European Union under the European Regional Development Fund grant no. POIR.04.04.00-00-2475/16-00. The project on range monitoring with J-PET detector is funded by the National Centre for Research and Development (NCBiR), grant no. LIDER/26/0157/L-8/16/NCBR/2017. MG and JB acknowledge the support within InterDokMed programme, project no. POWR.03.02.00-00-I013/16. PM acknowledges Foundation for Polish Science for support within TEAM programme, project no. TEAM/2017-4/39. We acknowledge the support of NVIDIA Corporation with the donation of the GPU used for this research. AR acknowledges Prof. Reinhard Schulte from Loma Linda University, CA, USA for mentoring and support.
 
%uncomment the following lines to place a figure
%\begin{figure}[htb]
%\centerline{%
%\includegraphics[width=12.5cm]{Fig1}}
%\caption{Plot of ...}
%\label{Fig:F2H}
%\end{figure}
%\section*{References}
\label{sec:references}
\bibliographystyle{ama}
\bibliography{references}

\end{document}